# Machine Learning-Driven Design of Mixed-Pitch Grating Couplers for Co-Packaged Optics Applications

Yu Dian Lim[1, *], Yun Da Chua[2], Wai Cheung Ma[2], Yeow Kheng Lim[2], Chuan Seng Tan[1, *]

[1]School of Electrical and Electronic Engineering, Nanyang Technological University

[1]50 Nanyang Ave, Singapore 639798

[2]Singapore Hybrid Integrated Next Generation μ-Electronics (SHINE) Centre, National University of Singapore

[2]5 Engineering Dr 1, Singapore 117608

*yudian.lim@ntu.edu.sg, tancs@ntu.edu.sg

## ABSTRACT

A mixed-pitch grating coupler which can couple a wide range of wavelengths is preferred in its application in co-packaged optics (CPO). However, the design and optimization of such grating coupler is complex. In this work, we developed software with integrated deep neural network (DNN) model to automatically design the mixed-pitch grating coupler from user-specified peak wavelengths and full-width half-maximum (FWHM) values. We first trained the DNN model with 10,000 rows of grating parameters-power spectrum datasets, where the power spectrum was simulated using finite-difference time domain (FDTD) technique. Upon training, we tested the model using ~1,000 different combinations of peak wavelengths and FWHM values. Among the combinations, 822 attempts have <15% error, while 351 attempts have <5% error when comparing the user-specified and FDTD-verified spectrum. Meanwhile, comparing the user-specified and FDTD-verified peak wavelengths, 844 attempts have peak wavelengths with absolute error (AE) < 0.002 µm. For FWHMs, 738 attempts have FWHM values with AE < 0.010 µm. We have also developed a graphical-user interface (GUI) to ease the usage of this software.

## INTRODUCTION

In the artificial intelligence era, the exponential surge in hyper-scale workloads has pushed traditional copper-based electrical interconnects to their physical limits, especially in data center bandwidth, power consumption, and latency [1]. To resolve this issue, Co-Packaged Optics (CPO) was introduced to integrate the photonics transceivers directly onto a single substrate alongside with high-performance ASICs, GPUs, or switches[2], [3].

CPO often involves integration of photonics channels with different wavelengths and spectral responses into the same platform. Thus, the components in each channel, including the grating couplers should be tailor-made for a specific wavelength or a specified range of wavelengths, depending on the designer's requirements[4].

Fundamentally, the pitch ($\Lambda$) in the grating coupler is directly corresponding to its operational wavelength ($\lambda_0$) described by the following equation:

$$\Lambda = \frac{\lambda_0}{n_{eff} - n_{clad}\sin\theta}$$

where $n_{eff}$, $n_{clad}$, and $\theta$ represent the effective index, cladding index, and the incident angle, respectively [5]. However, in the context of CPO, sharing of single input source of light by multiple channels is common. In such scenarios, a mixed pitch grating is often needed for multi-wavelength coupling [6], [7].

Designing mixed pitch grating for wide bandwidth, multi-wavelength coupling is complex, involving repeated attempts of finite-difference time-domain (FDTD) simulation. Thus, in this work, we introduce a machine learning-driven technique to design the mixed-pitch grating couplers for various wavelength requirements.

## METHODOLOGY

Figure 1 shows the schematic diagram of the mixed pitch gratings. As shown, the grating comprises of 4 regions, where all regions have different pitches (p) and duty cycles (dc) to couple wider range of wavelengths. Each region has 6 protrusions, regardless of the p and dc. The p and dc of region 1, 2, 3, 4 shall be named as p1, dc1, p2, dc2, p3, dc3, p4, and dc4.

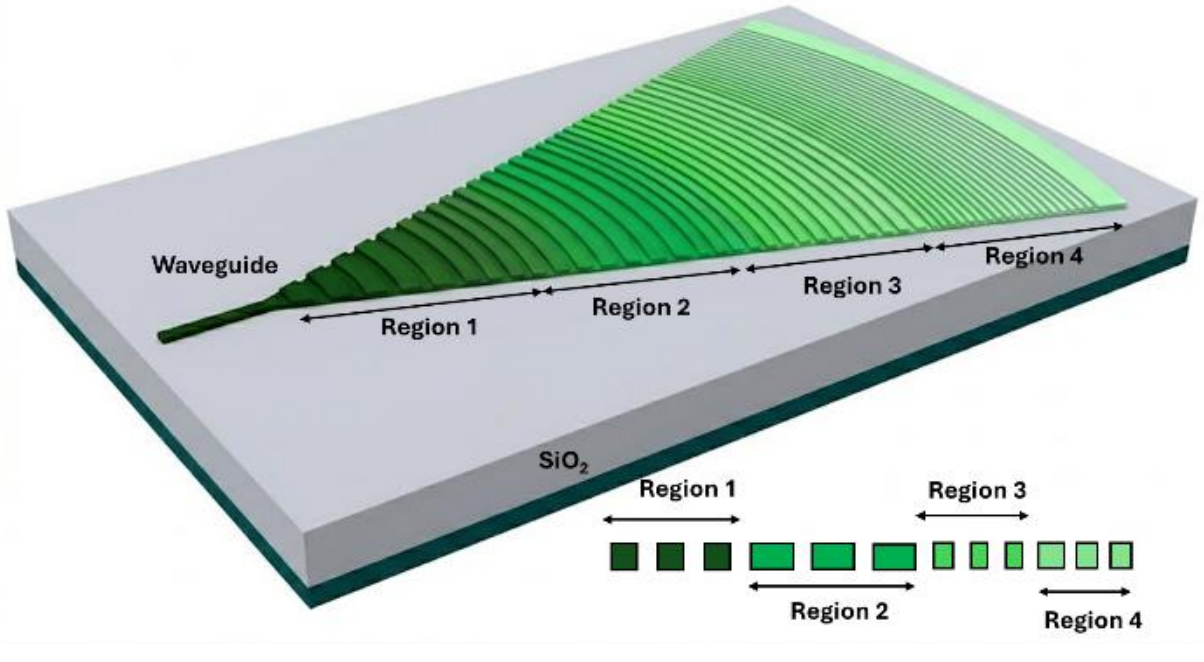

FIGURE 1. SCHEMATIC DIAGRAM OF THE MIXED PITCH GRATING

Figure 2 shows the software architecture of the machine learning-driven technique. To train the model, we first prepared ~10,000 sets of training data, comprising of ~10,000 different combinations of p1, dc1, p2, dc2, p3, dc3, p4, dc4 (grating parameters) and their corresponding FDTD-simulated spectrum. The model used in this work is deep neural network (DNN) model with 5 hidden layers, where each hidden layer has 256, 128, 64, 32, 16 nodes, respectively. The input of the model is the FDTD-simulated spectrum with 300 values; while the output of the model is the 8 grating parameters (p and dc) described earlier.

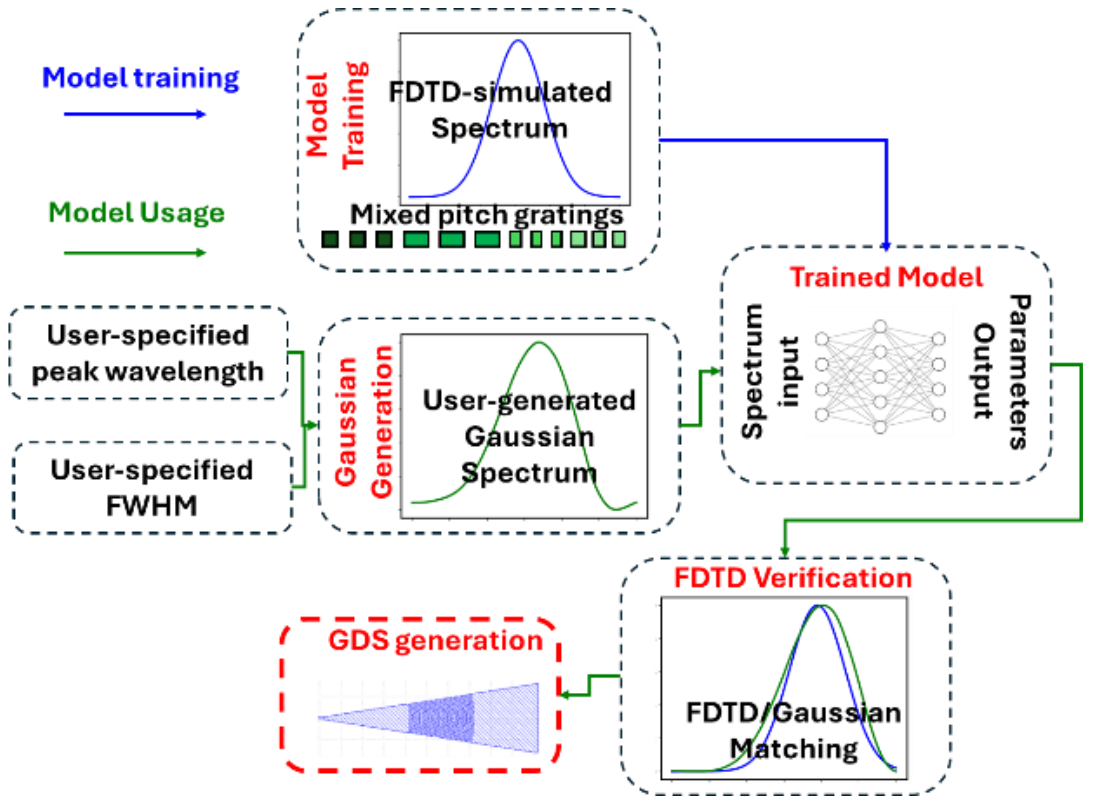

FIGURE 2. SOFTWARE ARCHITECTURE

After training the model, the trained model can be used to custom design the mixed-pitch grating couplers. First, users can specify the desired peak wavelength and full-width half-maximum (FWHM) values. Based on these two values, a Gaussian spectrum with 300 values will be generated. The Gaussian spectrum will be used as the input for the trained DNN model to suggest the corresponding grating parameters. The grating parameters are then used in FDTD simulation, to simulate the power spectrum of the fiber-to-grating coupling. The simulated spectrum will be normalized and matched with the user-generated Gaussian spectrum. If the user is satisfied with the match, the Graphic Data System (GDS) layout will be automatically drawn.

## RESULTS AND DISCUSSION

To test the model performance, we used ~1,000 combinations of FWHM = 0.08 – 0.145 µm and peak wavelengths = 1.535 – 1.615 µm as the inputs for the trained DNN model. The mean absolute error (MAE) values between the Gaussian and FDTD-simulated curves are computed. In general, the model performs better at higher peak wavelengths between 1.57 to 1.6 µm. and FWHM between 0.95 to 0.115 µm. The MAE values of this region ranged from 0.006 to 0.04. Since the spectrum were normalized to 0 – 1, this translates to 0.6 to 4% error. To better visualize the outcomes, we have labeled the "Region with <5% error" and Region with <15% error" in Figure 3(a). Among the ~1,000 combinations, 822 attempts have <15% error, while 351 attempts have <5% error.

The MAE results presented in Figure 3(a) were computed from the full spectrum of the FDTD-simulated spectrum and the user-defined Gaussian spectrum. However, in the context of fiber-to-grating coupling, the priority should be the peak wavelength and the FWHM of the spectrum. Figure 3(b) shows the absolute error (AE) between the peak wavelengths of the FDTD-simulated and user-defined Gaussian spectrum. As shown, there are substantial instances where AE = 0 µm, which means that the user-defined peak wavelength matches exactly with the peak wavelength of FDTD-simulated spectrum. Within the ~1,000 combinations, there are significant number of instances where the peak wavelength's AE is < 0.002 µm, where most of the instances shows AE is < 0.010 µm. Among the ~1,000 combinations, 844 attempts have peak wavelengths with AE < 0.002 µm. For FWHMs, most instances show AE < 0.010 µm, with significant number of instances where AE < 0.005 µm (Figure 3(b)). Among the ~1,000 combinations, 738 attempts have FWHM values with AE < 0.010 µm.

For further investigation, we attempted to experimentally verify the spectrum produced by AI-generated grating couplers. The GDS layout produced by the software (Figure 2) was fabricated on a 220nm silicon-on-insulator (SOI) platform where the SOI layer is partially etched for 80 nm, followed by depositing 2 µm of $SiO_2$ layer onto the SOI. The obtained sample chip is then measured using a typical top-coupling optical measurement setup shown in Figure 4(a). The measurements are done on grating-waveguide-grating structures shown in Figure 4(b) under 1.45 – 1.65 µm wavelengths, The details of the measurement were described in our previous work [7], [8].

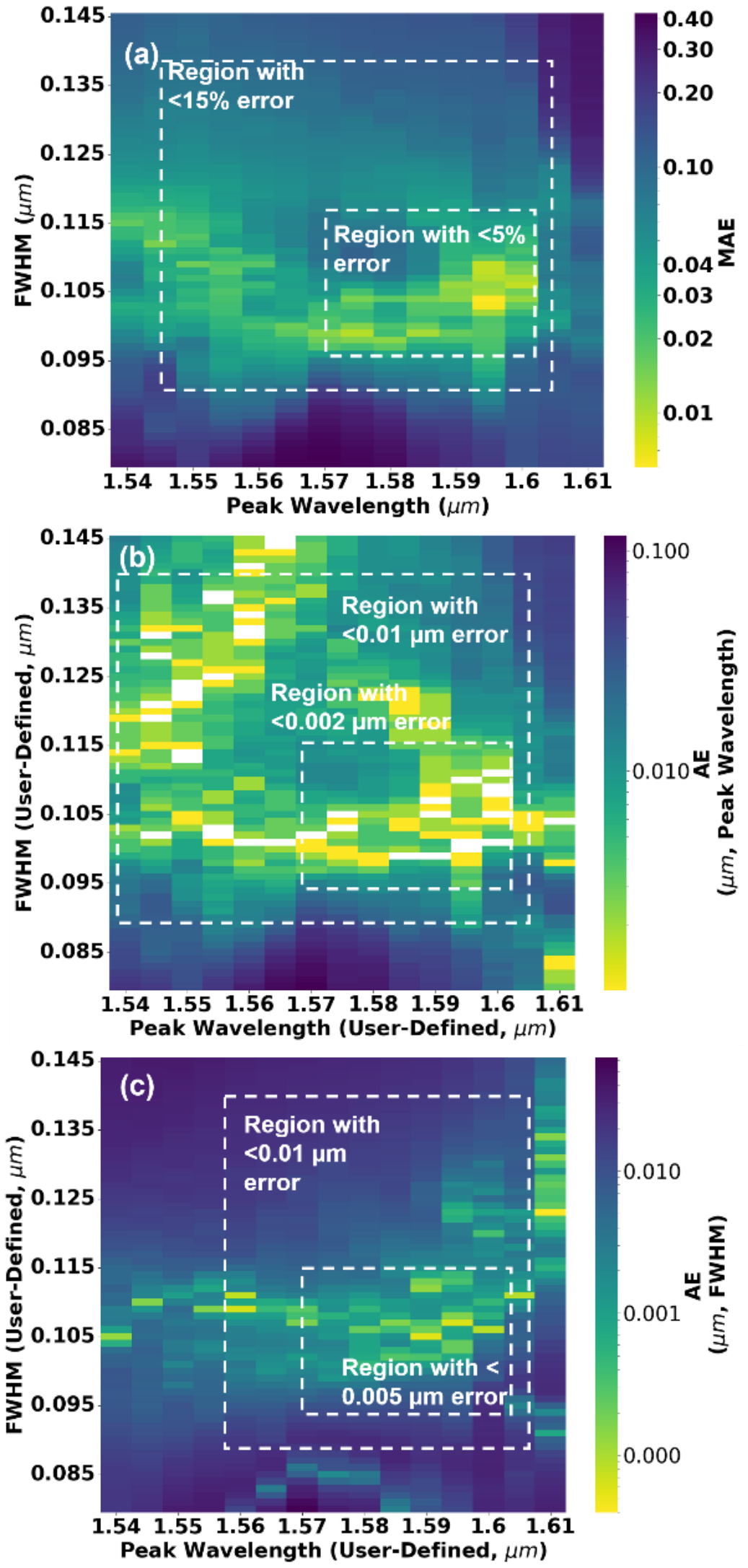

FIGURE 3. MODEL PERFORMANCE IN FWHM = 0.08 – 0.145 µM AND PEAK WAVELENGTHS = 1.535 – 1.615 µM. (a). MAE BETWEEN FDTD-SIMULATED SPECTRUM AND USER-DEFINED GAUSSIAN SPECTRUM, (b). ABSOLUTE ERROR BETWEEN PEAK WAVELENGTHS OF FDTD-SIMULATED SPECTRUM AND USER-DEFINED SPECTRUM, (c) ABSOLUTE ERROR BETWEEN FWHMS OF FDTD-SIMULATED SPECTRUM AND USER-DEFINED SPECTRUM

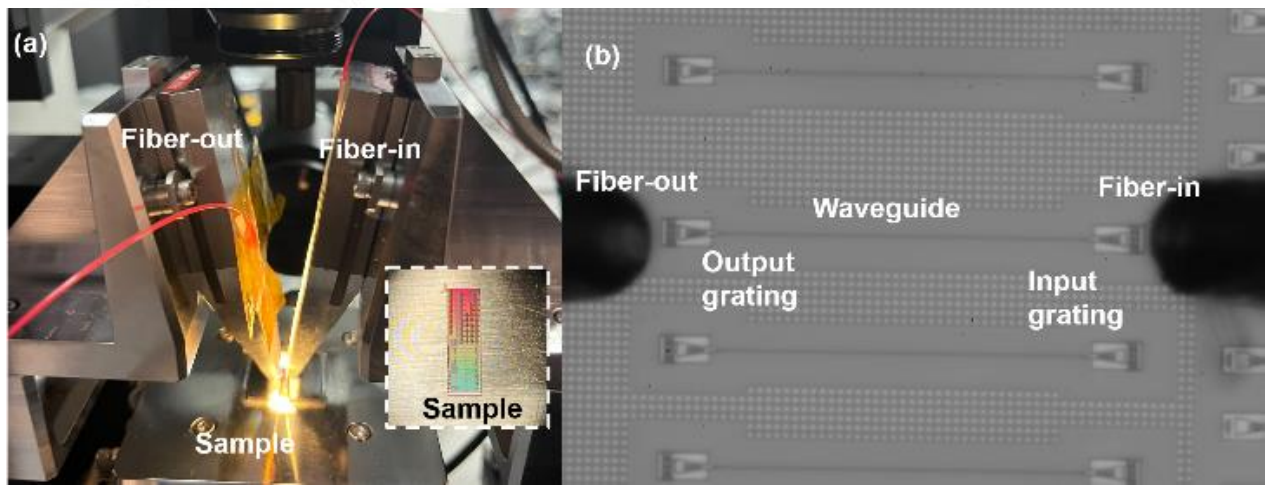

FIGURE 4. (a) EXPERIMENTAL SETUP, (b) MICROSCOPIC IMAGE OF THE GRATING-WAVEGUIDE-GRATING STRUCTURE

Figure 5 shows 3 sets of FDTD-simulated spectrum, user-defined Gaussian spectrum, and experimentally measured spectrum. Comparing across the simulated, Gaussian, and measured spectrum, the AE values of the peak wavelengths are generally < 0.005 µm. The FWHM values show a larger deviation as compared to peak wavelengths. The deviation of FWHM is especially prominent when comparing between FDTD-simulated and experimentally measured spectrum. This can be attributed to several factors. First, the 10,000 grating parameters-spectrum pair used for model training (ref. Figure 2). To reduce the simulation time, the FDTD simulation was done using a two-dimensional (2D) model. However, the grating couplers were fabricated and measured in three-dimensional (3D) manner (ref. Figure 1). In 2D model, all 4 regions illustrated in Figure 1 are treated equally. However, in 3D reality, all 4 regions have different sizes. For instance, region 4 is significantly smaller than region 1 due to the taper shape nature of the grating coupler. Apart from that, manufacturing errors, such as errors when etching the grating structure, may affect the experimental measurement results of the grating couplers [9], [10], [11].

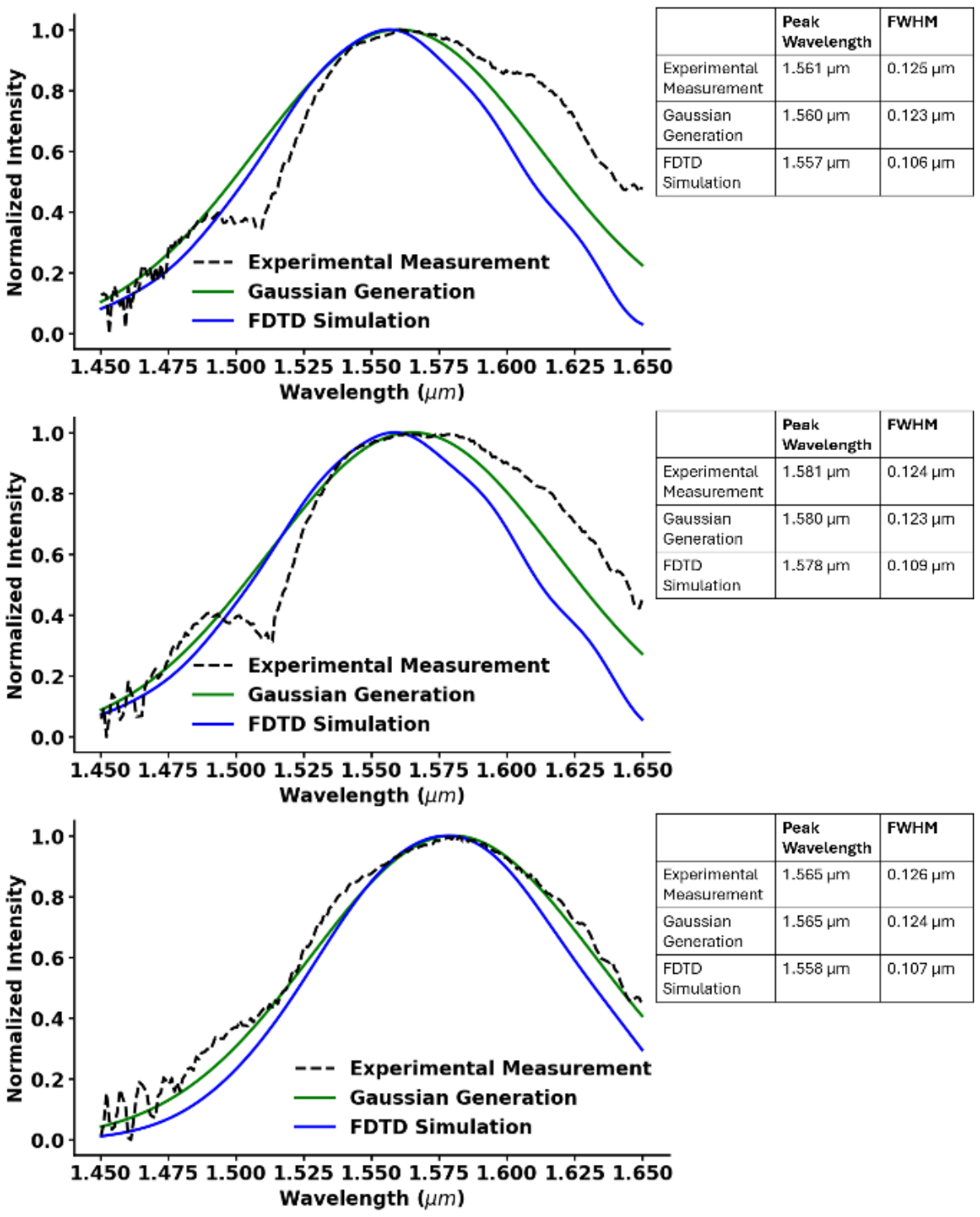

FIGURE 5. COMPARISONS BETWEEN FDTD-SIMULATED SPECTRUM, USER-DEFINED GAUSSIAN SPECTRUM, AND EXPERIMENTALLY MEASURED SPECTRUM

Despite the deviations in FWHM values between FDTD-simulated spectrum and experimentally measured spectrum, the morphology of the curves obtained from three separated sources are generally similar, as shown in Figure 5. Nevertheless, based on the MAE and AE results shown in Figure 3, it can be preliminarily deduced that the trained DNN model can accurately "suggest" a suitable set of grating parameters (p1, dc1, p2, dc2, p3, dc3, p4, dc4) to match the desired peak wavelength and FWHM. Further improvements can be made. For instance, the FDTD simulation can be improved to better match the experimental measurements, which requires a mass amount of experimental measurement data. Apart from that, the DNN model can be trained with a mixture of experimental and simulation data to enhance the accuracy of the model.

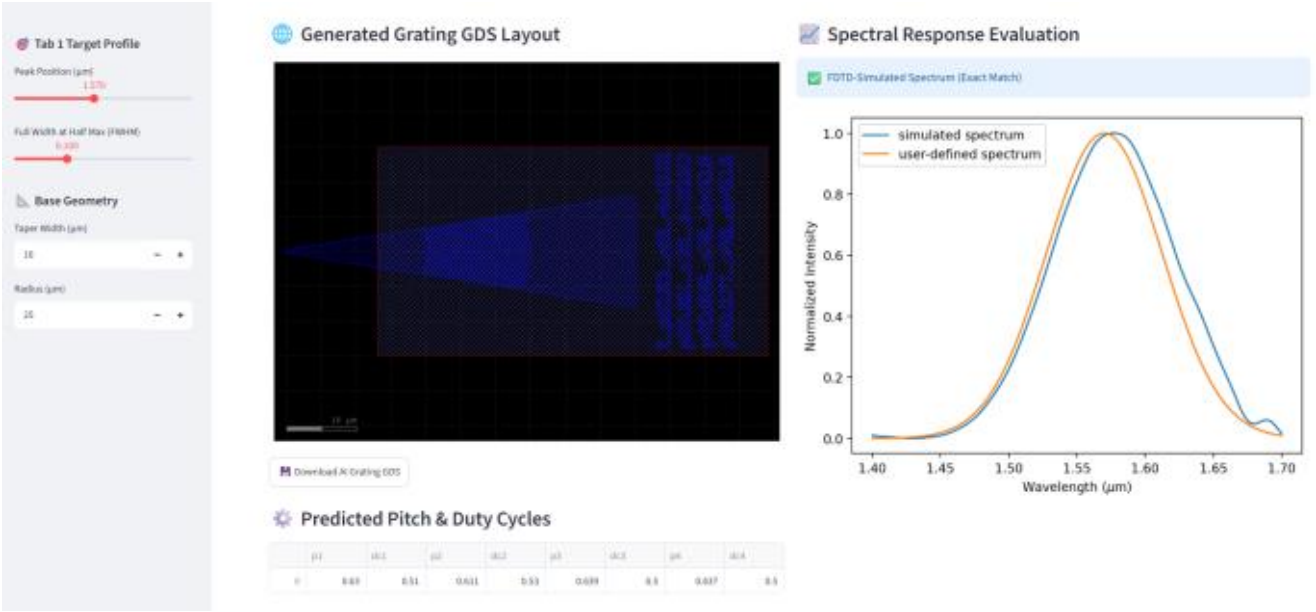

FIGURE 6. GUI OF THE DEPLOYED MODEL

Apart from the abovementioned studies, we have deployed the model into software where the architecture structure was described in Figure 2. To ease the usage, we have prepared a graphical-user interface (GUI) using streamlit. The GUI is shown in Figure 6. Users can access this GUI in www.inphai.com to test this model. At the same time, users can download the desired GDS layout suggested and designed by the model. The demonstration video of the deployed model is shown in ref. [12].

## CONCLUSION

In this work, we have introduced a 4-region mixed-pitch grating coupler for possible applications in co-packaged optics (CPO). The mixed pitch grating coupler can couple a range of wavelengths, which enables multiple channels of the transceiver units in the CPO module to share a single laser light input. To better design the mixed pitch grating couplers, we have developed software architecture with integrated deep neural network (DNN) model. Users can specify the desired peak wavelength and the FWHM from the power spectra by the grating coupler, the software will suggest corresponding grating parameters. The grating parameters will first be subjected to FDTD simulation for verification, then, the GDS layout of the grating coupler will be generated.

We conducted ~1,000 attempts to test the trained DNN model with various combinations of peak wavelengths and FWHMs. Comparing the normalized spectrum, 822 attempts have <15% error, while 351 attempts have <5% error. Meanwhile, comparing the user-specified and FDTD-verified peak wavelengths, 844 attempts have peak wavelengths with AE < 0.002 µm. For FWHMs, 738 attempts have FWHM values with AE < 0.010 µm. To enable wider range of users, we have developed a GUI and deployed the software to www.inphai.com where users can access and test the model. Users can also download the desired GDS layout according to their preferred peak wavelengths and FWHM values.

## ACKNOWLEDGEMENT

This research/project is supported by A*STAR under the Semiconductor RIE Flagship Programme: Singapore Hybrid-Integrated Next-Generation μ-Electronics (SHINE) Centre 2.0 (H26-MSE1554)..This work was supported by Ministry of Education Tier-1 RG135/23, RG67/25, and RT3/23; National Research Foundation (NCAIP(NRF-MSG-2023-0002). The chip in this work was taped out through GlobalFoundries (formerly AMF) under its multi-project wafer (MPW) services for passive devices. The preparation of Python codes used in this work was assisted by open-source large language model (LLM) tool, Gemini. At the same time, Gemini was also used to assist in the preparation of Figure 1.